\documentstyle[12pt,epsfig]{article}
\def\zz{\vspace*{-1.3mm}}

\def\be{\begin{equation}}
\def\ee{\end{equation}}
\def\bea{\begin{eqnarray}}
\def\eea{\end{eqnarray}}

\begin{document}
\begin{center}
{\Large \bf A possible  alternative to the Breit frame in DIS }\\

\vspace*{1cm}
{ \bf Boris B. Levtchenko}\\
\vspace*{0.5cm}
{Institute of Nuclear Physics, Moscow State University,\\
119899 Moscow, Russia}

\date{}

\vspace*{0.5cm}
 
{\bf Abstract}\\

A new Lorentz frame  for DIS jet finding is suggested.
\end{center}

\vspace*{4.5cm}
\begin{center}

Talk  presented at the 6th Int. Workshop \\
  on Deep Inelastic Scattering and QCD: DIS 98\\
   (Brussels, 4-8 April 1998)
\end{center}

\newpage

\section{\bf  Motivation}
The traditional choice  of a Lorentz frame to perform jet finding  
in $e^{\pm}p$ DIS final
state is the Breit frame, since \cite{Webber} in such  a frame
$k_{\bot}$-type jet clustering  algorithms  would preserve factorization
 which is
 an important feature of QCD. 
Vectors, whether partons or calorimeter cells,
used as the input  to the jet algorithm,
 have to be boosted from the  laboratory frame to the
Breit frame and then clustered into jets. However, boosting from the laboratory
frame to the  Breit frame introduces systematic errors \cite{Thompson}
 that may affect
the jet-finding results. In particular \cite{Magill}, when boosting
calorimeter cells, a problem arises near the outer edges of the forward  and
rear sections of a cylindrical calorimeter system.  This is the region where
the cells are least projective radially, 
and the longitudinal variation in the energy deposit in these cells results
in large differences in the polar angle between cells after boosting
them to the Breit frame. Many methods have been tried \cite{Magill} to
 reduce this problem,
but in the end, none give a satisfactory result\footnote{This may be 
one of the reasons why there are still  very few published results  with data
analysis from HERA using the $k_{\bot}$ jet algorithm.}.

Our goal is to demonstrate the existence of a new Lorentz frame, more suitable
for DIS jet finding.

\section{\bf  The Breit Frame}
In order to define fully a jet
clustering algorithm one needs to introduce \cite{Webber} an
auxiliary vector $\bar{p}$ of the form 
$$\bar{p}\,=\,x\,f(Q^2){\cal P}\,+\,g(Q^2)q\,.$$
Here ${\cal P}$ and $q$ are the incoming proton and virtual photon 
four-momenta and $f,\
g$ are any function of $Q^2$. The simplest example of a suitable auxiliary
vector is 
$$ \bar{p}\,=\,2x\,{\cal P}\,+\,q$$ 
with $ \bar{p}^2\,\simeq\,Q^2$ and  $ \bar{p}\cdot q\,=\,0$.
The last equation  can be used to specify a frame of reference in which
the cluster resolution variables $ d_{ij}$ are to be evaluated.
For frames of reference where a virtual photon is purely space-like
 ($q^*_0=0$) there are two
solutions of the equation $\vec{\bar{p^*}}\cdot\vec{q^*}\,=\,0$. The first
 solution
($\vec{\bar{p^*}}\,=\,0$) corresponds to the rest frame of $\bar{p}$ and known 
as the Breit frame (BF) of reference.

In terms of $\bar{p}$ the Lorentz parameters of the BF are as
follows
\be
 \gamma\,=\,\bar{p}_o/\sqrt{\bar{p}^2},\ \ \ \
\vec{\eta}\,=\,\vec{\bar{p}}/\sqrt{\bar{p}^2}.
\ee
Fig.\,1a shows the Lorentz factor $\gamma$ as a function of $x$ at different
$y$. The arrow shows a unique point  $x=x_o=k/P$, $y=1$ where the HERA
laboratory and  Breit frames coincide ($\gamma=1$)\,.
Here $k$ and $P$ are the incoming lepton and proton momenta.
A large variation of $\gamma$ with ($x,\,Q^2$)   causes problems noted in 
Sec.\,1 and discussed in \cite{Thompson}.

\newpage

\vspace*{-3.4cm}
\hspace{-0.5cm}
\begin{minipage}[t]{7.5cm}
\setlength{\unitlength}{\textwidth}
\begin{picture} (1.0,1.35) (0.1,0)
\mbox{\epsfig{file=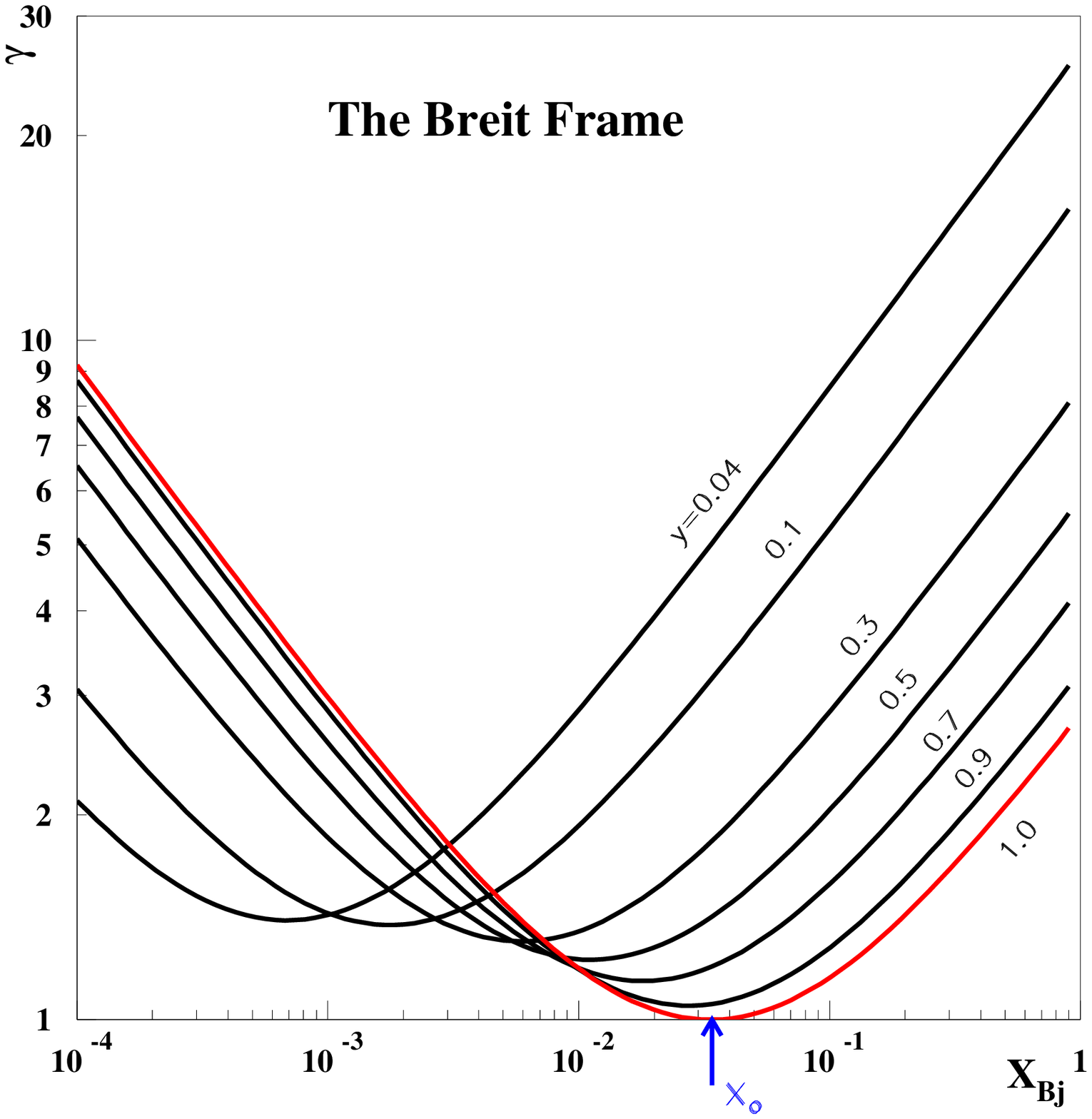,width=1.2\textwidth,height=1.20\textwidth}}
\end{picture}

\end{minipage}
\begin{minipage}[t]{7.5cm}
\begin{picture} (1.0,1.35) (-0.3,0)
\mbox{\epsfig{file=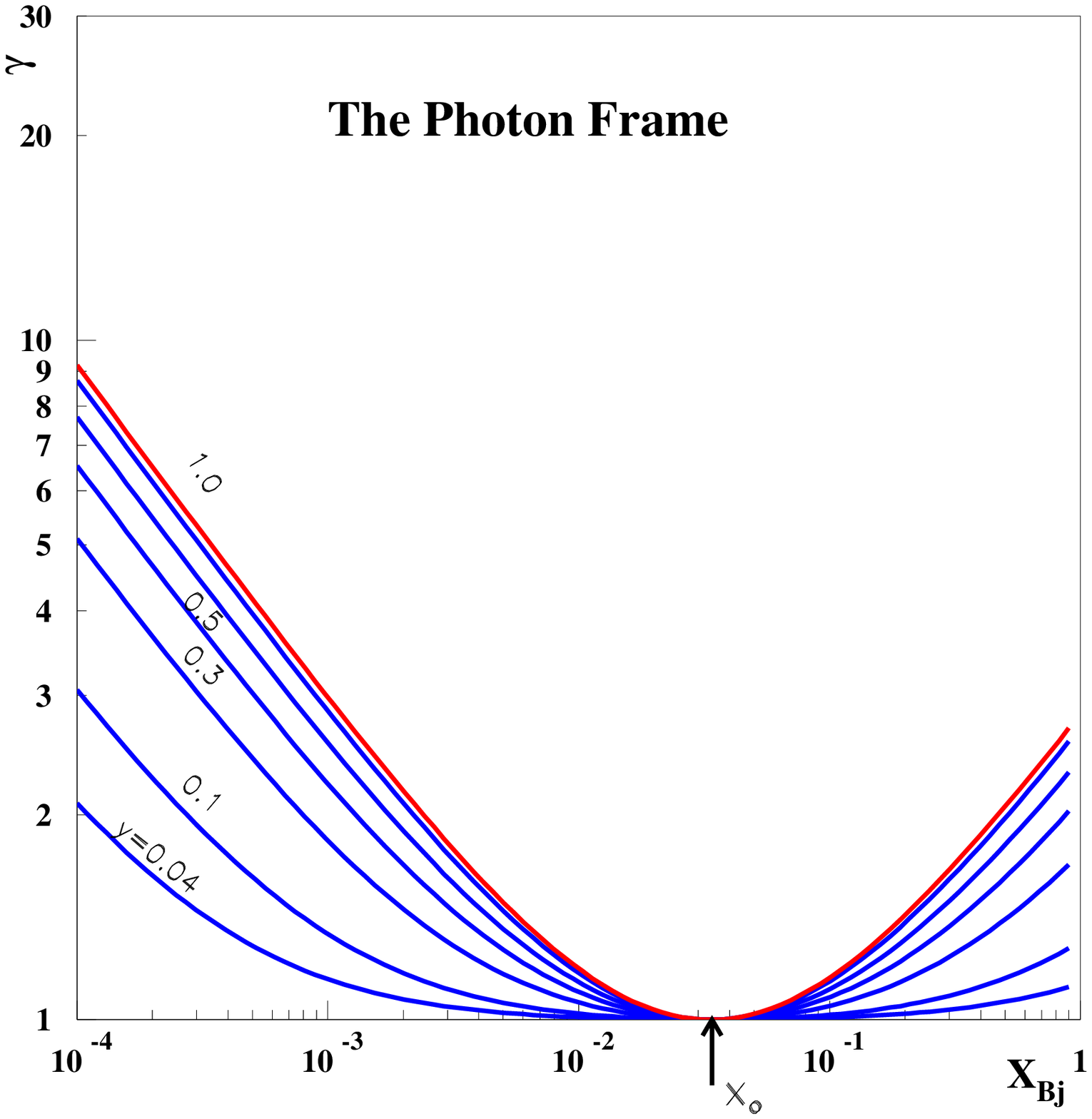,width=1.2\textwidth,height=1.20\textwidth}}
\end{picture}
\end{minipage}

\vspace*{-1mm}
\begin{center}
{\bf Figure 1}
\end{center}

\section{\bf  The Photon Frame}
The second solution of the equation
$$ \bar{p^*_o}q_o^*\,-\,\vec{\bar{p^*}}\cdot\vec{q^*}=0$$ 
corresponds to $\vec{\bar{p^*}}\neq 0$ and $\vec{\bar{p^*}}\,\bot\, \vec{q^*}$
at $q^*_o\,=\,0$.
In the general form the Lorentz parameters of a new frame, called 
the Photon frame of reference, are
\be
\gamma\,=\,\ell_o/\sqrt{\ell^2}\,,\hspace{0.8cm}\vec{\eta}\,=\,
\vec{\ell}/\sqrt{\ell^2}
\vspace*{-2mm}
\ee 
 with 
$$\ell=(\sqrt{q^2_o\,+\,Q^2}\,,\,q_o\vec{q}/\sqrt{\vec{q^2}})$$ 
and $ \ell^2=Q^2\,,\ \ell\cdot q=0\,.$

Here we would like to enumerate some properties of the new frame.
From (2) one sees that the lababoratory and Photon frames are connected by
 a boost along the 
direction of the momentum transfer vector $\vec{q}$.
Fig.\,1b shows the Lorentz factor (2) as a function of $x$ at different $y$.  
At $x>10^{-3}$\ $\gamma_{Ph}$ 
depends on ($x,\,Q^2$) values in a very different way compared with 
 $\gamma_{Br}$
in Fig.\,1a. In the range $10^{-2}\,< x\,<\,10^{-1}$ the Photon frame
(PF) is 
very close to the HERA frame though  $Q^2$ varies significantly.
At $x=x_o$ (the arrow in Fig.\,1b) the PF  coincides with the HERA
frame along the line in the phase space independent of $Q^2$ and $y$ values
 and a virtual photon is pure space-like in the laboratory frame 
of reference.

Deep inelastic lepton-nucleon scattering in the PF is described 
in the  parton model (zeroth order QCD) by the space diagram in Fig.\,2a. 
An auxiliary angle between the scattered lepton and quark is denoted as
 $\alpha$.
Angles $\delta ,\,\theta ,\,\xi\,$ and  $\alpha$ 
relate to $q_o$, $Q^2$ and the incoming lepton and proton energies, 
$\epsilon$, $E$, as follows

\newpage

\vspace*{-1mm}
\hspace{-1.8cm}
\begin{minipage}[t]{7.5cm}
\setlength{\unitlength}{\textwidth}
\begin{picture} (1.0,1.05) (0.1,0)
\mbox{\epsfig{file=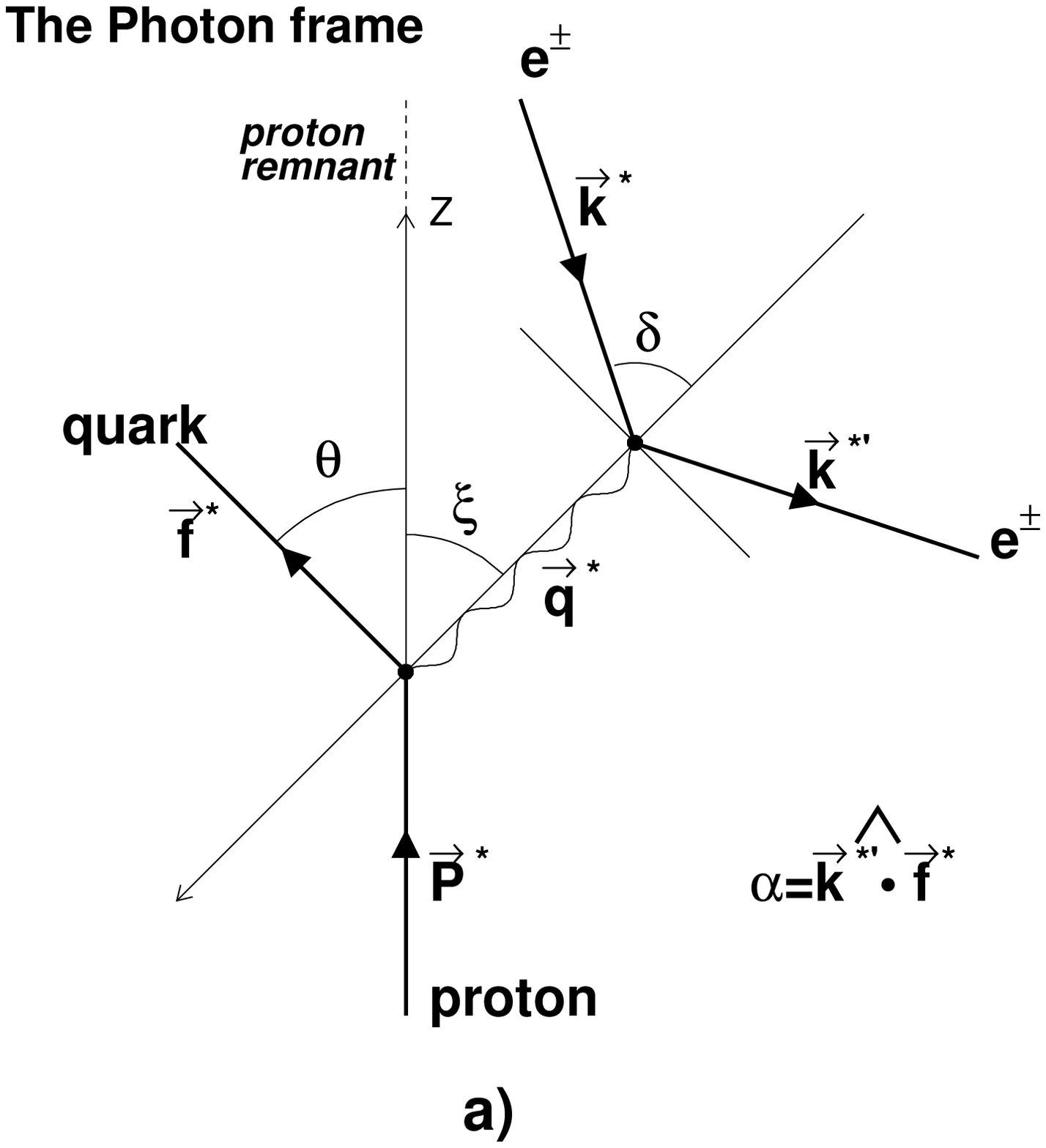,width=1.20\textwidth,height=1.20\textwidth}}     
\end{picture}
\end{minipage}\hspace{-1.4cm}
\begin{minipage}[t]{7.5cm}
\setlength{\unitlength}{\textwidth}
\begin{picture} (1.0,1.05) (-0.3,0)
\mbox{\epsfig{file=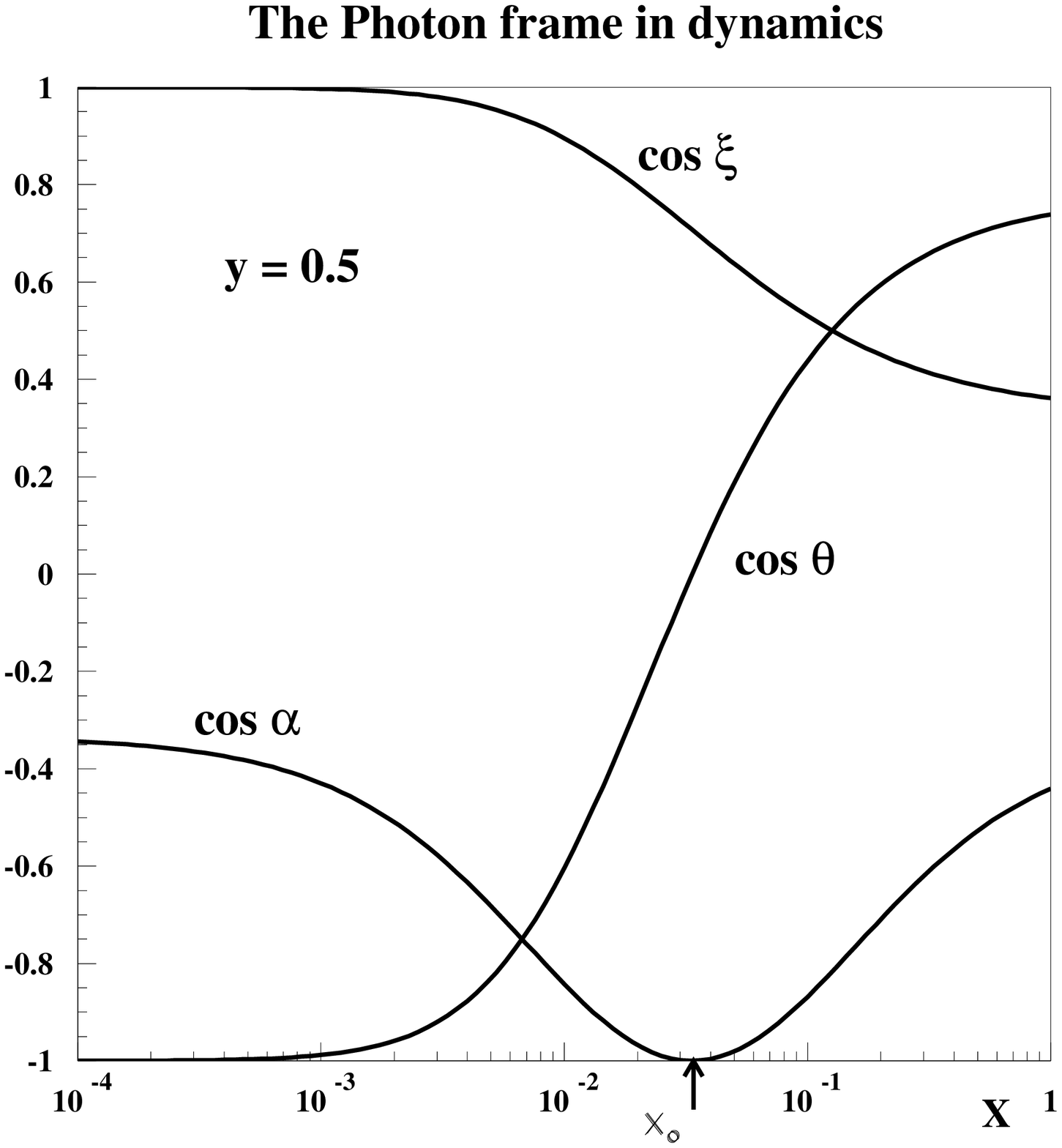,width=1.20\textwidth,height=1.20\textwidth}}
\end{picture}
\end{minipage}

\vspace*{-1mm}
\begin{center}
{\bf Figure 2}
\end{center}

\vspace*{2mm}
$$cos\,\delta=\frac{\sqrt{Q^2\,+\,q^2_o}}{2\epsilon\,-\,q_o}\,,
\ \ \ cos\,\xi=\frac{\sqrt{Q^2\,+\,q^2_o}}{2xE\,+\,q_o}\,,\eqno(3)
$$
  
$$cos\,\theta=1\,-\,2\,cos^2\xi\,, 
\ \ cos\,\alpha=1\,-\,\frac{2}{y}cos\,\delta\cdot cos\,\xi\eqno(4)
$$ 
\noindent with
$$ q_0=\frac{(k-xP)Q^2}{2x(kE+\epsilon P)}\simeq(k-xP)y\,.
$$

%
\noindent Due to the relations (4) in between $\delta ,\,\theta ,\,\xi\,$ 
and  $\alpha$ there are only two independent angles.
Fig.\,2b shows variation of these angles with $x$ at $y=0.5$. 
 At $x< 10^{-3}$ the BF and the PF are very close to
each other ($\theta\simeq\pi,\ \xi\simeq0$).
Direct comparison of  (1) and (2) also confirms the last conclusion,
since at small $x$ 
 $\gamma_{Ph}\,\sim\,\gamma_{Br}\,\simeq\,q_o/Q\,,\hspace{0.4cm}
\vec{\eta}_{Ph}\,\sim\,\vec{\eta}_{Br}\,\simeq\,\vec{q}/Q$.
In the parton model the line $x=x_o$ has a special significance, since 
both incoming and outgoing  $e^{\pm}$ and parton have the same energy and
back-scatter off each other ($\alpha=\pi$).

We point out  that jet finding in the PF
preserves factorization. Careful analysis \cite{Frames} of  examples given in
\cite{Webber} shows that in term of the  vector  
$\bar{p^*}=2x{\cal P}^*+q^*$  in the PF 
the $k_{\bot}$-type resolution variable $b_{ij}$  has the same form 
as Eqs. (25)-(26) of Ref. \cite{Webber}.
As compared with the PF the BF current hemisphere
due to the static geometry  
 is dominated by the fragments of 
the struck quark. 
This makes comparisons of multiplicities in $e^{+}e^{-}$ and 
the current region of $e^{\pm}p$  easier in the BF. 
On the other hand,
to perform the DIS jet finding in $e^{\pm}p$ it is preferable
to use the PF because it reduces the above-mentioned problems.

\vspace*{+2mm}
\section{\bf Conclusions} 
A new Lorentz frame, called the Photon frame,
with
a pure space-like virtual photon is found.
Many features of the Photon frame make of it attractive for jet finding.
In the kinematical region interesting for DIS jet study  
boosts are small,   substantially reducing the  
systematic errors.

\vspace{3mm}
\noindent{\bf Acknowledgment.} I would like to thank the DESY Directorate
and the  organisers of this meeting for financial support and to acknowledge 
friendly assistance and discussions with E. De Wolf, J. Hartmann, R. Klanner, G.
Wolf. I am grateful to P. Bussey for reading the manuscript and comments.
 This work supported in part under  the DFG grant \#436 RUS 113/248/1.

%

\vspace{2mm}

\begin{thebibliography}{99}
\bibitem{Webber}
B.R.~Webber, J.Phys. G: Nucl.~Part.~Phys. {\bf 19}(1993)\,1567
\bibitem{Thompson}
G.Thompson et.al, J.Phys.G: Nucl.Part.Phys. {\bf 19}(1993)1575
\bibitem{Magill}
S.~Magill,  A measurement of the inclusive jet production cross section\zz
in DIS at HERA, The ZEUS Collab., Internal note 94-154;\ S.~Magill,\zz
 B.~Musgrave, T.~Trefzger, Di-jet rates in DIS using the $k_{\bot}$ algorithm,\zz
 The  ZEUS Collab., Internal note 95-029
\bibitem{Frames}
B.~B. Levtchenko,  in preparation

\vspace*{-2mm}
\end{thebibliography}
\end{document}